\documentclass[10pt,twoside]{article}
\usepackage{amsmath,amsfonts,amssymb}
\usepackage{latexsym}
\usepackage{dcolumn}
\usepackage{graphicx,epsfig}
\usepackage{amsthm}
\usepackage{hyperref}

\begin{document}

\begin{flushright}
{\tt IISER(Kolkata)/GR-QC\\ \today}
\end{flushright}
\vspace{1.5cm}

\begin{center}
{\Large \bf Effects of the Modified Uncertainty Principle on the Inflation Parameters}
\vglue 0.5cm
Barun Majumder\footnote{barunbasanta@iiserkol.ac.in}
\vglue 0.6cm
{\small {\it Department of Physical Sciences,\\Indian Institute of Science Education and Research (Kolkata),\\
Mohanpur, Nadia, West Bengal, Pin 741252,\\India}}
\end{center}
\vspace{.1cm}

\begin{abstract} 
In this Letter we study the effects of the Modified Uncertainty Principle as proposed in \cite{b1} on the inflationary dynamics of the early universe in both standard and Randall-Sundrum type II scenarios. We find that the quantum gravitational effect increase the amplitude of density fluctuation, which is oscillatory in nature, with an increase in the tensor-to-scalar ratio.
\vspace{5mm}\newline Keywords: Generalized Uncertainty Principle, Quantum Gravity Phenomenology, Inflation
\end{abstract}
\vspace{.3cm}

The idea that the uncertainty principle could be affected by gravity was first given by Mead \cite{c1}. Later modified commutation relations between position and momenta commonly
known as Generalized Uncertainty Principle ( or GUP ) were given by candidate theories of quantum gravity ( String Theory, Doubly Special
Relativity ( or DSR ) Theory and Black Hole Physics ) with the prediction of a minimum measurable length \cite{b2,kem,b7}. Similar kind of
commutation relation can also be found in the context of Polymer Quantization in terms of Polymer Mass Scale \cite{c}. Recently authors in \cite{cnew} claimed the existence of minimal space and four-volumes but not minimal length.
\par
The authors in \cite{b1} proposed a Modified Uncertainty Principle (from now we denote this as MUP to distinguish it from the Generalised Uncertainty Principle introduced in \cite{kem}) which is consistent with DSR theory, String theory and Black Hole Physics and which says
\begin{equation}
\left[x_i,x_j\right] = \left[p_i,p_j\right] = 0 ,
\end{equation}
\begin{equation}
\label{g2}
[x_i, p_j] = i \hbar \left[  \delta_{ij} -  l  \left( p \delta_{ij} +
\frac{p_i p_j}{p} \right) + l^2  \left( p^2 \delta_{ij}  + 3 p_{i} p_{j} \right)  \right],
\end{equation}
\begin{align}
\label{g3}
 \Delta x \Delta p &\geq \frac{\hbar}{2} \left[ 1 - 2 l <p> + 4 l^2 <p^2> \right]  \nonumber \\
& \geq \frac{\hbar}{2} \left[ 1  +  \left(\frac{l}{\sqrt{\langle p^2 \rangle}} + 4 l^2  \right)  \Delta p^2  +  4 l^2 \langle p \rangle^2 -  2 l \sqrt{\langle p^2 \rangle} \right],
\end{align}
where $ l=\frac{l_0 l_{pl}}{\hbar} $. Here $ l_{pl} $ is the Plank length ($ \approx 10^{-35} m $). It is normally assumed that the dimensionless
parameter $l_0$ is of the order unity. If this is the case then the $l$ dependent terms are only important at or near the Plank
regime. But here we expect the existence of a new intermediate physical length scale of the order of $l \hbar = l_0 l_{pl}$. We also note
that this unobserved length scale cannot exceed the electroweak length scale \cite{b1} which implies $l_0 \leq 10^{17}$. These equations are
approximately covariant under DSR transformations but not Lorentz covariant \cite{b7}. These equations also imply
\begin{equation}
\Delta x \geq \left(\Delta x \right)_{min} \approx l_0\,l_{pl}
\end{equation}
and
\begin{equation}
\Delta p \leq \left(\Delta p \right)_{max} \approx \frac{M_{pl}c}{l_0}
\end{equation}
where $ M_{pl} $ is the Plank mass and $c$ is the velocity of light in vacuum. It can be shown that equation (\ref{g2}) is satisfied by the
following definitions $x_i=x_{oi}$ and $p_i=p_{oi} (1 - l\,p_o + 2\,l^2\,p_o^2)$, where $x_{oi}$, $p_{oj}$ satisfies $[x_{oi}, p_{oj}]= i \hbar \delta_{ij}$. Here we can interpret $p_{oi}$ as the momentum at low energies having the standard representation in position space ($ p_{oi}=-i\hbar \frac{\partial}{\partial x_{oi}}$) with $p_o^2=\sum_{i=1}^3 p_{oi}p_{oi}$ and $ p_i $ as the momentum at high energies. The effect of this proposed MUP is well studied recently for some well known physical systems in \cite{z16,cite5}.
\par
The authors in \cite{kem} derived a generalized uncertainty principle and studied the quantum-mechanical structure of the principle which describes the minimal length as a minimal uncertainty in position measurements. With a plausible type of ultra-violet cut off \cite{cite1} we can continuously generate co-moving modes during inflation. Later \cite{cite2,cite3,cite4} studied different aspects of inflation using the GUP. Recently the authors in \cite{cite5} examined the effects of the MUP in the scenario of post inflation preheating and showed that there can either be an increase or a decrease in parametric resonance and a corresponding change in particle production. Guided by these ideas in this short Letter we examine the effects of the MUP on the slow-roll inflation parameters in standard and 
Randall-Sundrum type II scenarios. Here we are going to compute the fluctuation spectrum with our MUP and compare the result with the spectrum calculated earlier in the Heisenberg setup and the GUP.
\par 
The slow-roll parameters for inflation are given by \cite{cite6}
\begin{equation}
\epsilon_V = \frac{M^{2}_{4}}{16 \pi}~\Big(\frac{V'}{V}\Big)^{2}~~,
\end{equation}
\begin{equation}
\eta_V = \frac{M^{2}_{4}}{8 \pi}~ \frac{V''}{V}~~,
\end{equation}
where $V(\phi)$ is the potential responsible for inflation and $M_4$ is the four-dimensional fundamental scale. For {\it new inflation} or {\it slow-roll inflation} we require
\begin{equation}
\frac{1}{2}\dot{\phi}^2 \ll V(\phi)
\end{equation}
and
\begin{equation}
3H\dot{\phi} \simeq -V'(\phi)~~.
\end{equation}
If we consider an inflating background with the metric $ds^{2}=dt^{2}-a\left( t\right) ^{2}dx^{2}$ and the scale factor $\ a\left( t\right)\sim e^{Ht}$ then the scalar field equation (or the Klein-Gordon equation) in this background is given by
\begin{equation}
\ddot{\phi}+3H\dot{\phi}-\nabla ^{2}\phi =0~~.
\end{equation}
With $\eta =-\frac{1}{aH}$ as the conformal time and redefining the field as $\mu =a\phi $ we get
\begin{equation}
\label{deq1}
\mu _{k}^{\prime \prime }+\left( k^{2}-\frac{a^{\prime \prime }}{a}\right) \mu _{k}=0
\end{equation}
in Fourier space. Here $k=ap$ and $p$ is the physical momentum which is red-shifting with the expansion. In terms of the time dependent oscillators we can write
\begin{eqnarray}
\mu _{k}\left( \eta \right)  &=&\frac{1}{\sqrt{2k}}\left[ a_{k}\left( \eta \right) +a_{-k}^{\dagger }\left( \eta \right) \right]    \\
\pi _{k}\left( \eta \right)  &=&-i\sqrt{\frac{k}{2}}\left[ a_{k}\left( \eta \right) -a_{-k}^{\dagger }\left( \eta \right) \right] ~~.
\end{eqnarray}
The oscillators can be expressed in terms of their values at some fixed time $\eta _{0}$,
\begin{eqnarray}
a_{k}\left( \eta \right)  &=&u_{k}\left( \eta \right) a_{k}\left( \eta
_{0}\right) +v_{k}\left( \eta \right) a_{-k}^{\dagger }\left( \eta
_{0}\right)    \\
a_{-k}^{\dagger }\left( \eta \right)  &=&u_{k}^{\ast }\left( \eta \right)
a_{-k}^{\dagger }\left( \eta _{0}\right) +v_{k}^{\ast }\left( \eta \right)
a_{k}\left( \eta _{0}\right)~~.  
\end{eqnarray}
Plugging this back into the earlier set of equations we get
\begin{eqnarray}
\mu _{k}\left( \eta \right)  &=&f_{k}\left( \eta \right) a_{k}\left( \eta
_{0}\right) +f_{k}^{\ast }\left( \eta \right) a_{-k}^{\dagger }\left( \eta
_{0}\right)    \\
\pi _{k}\left( \eta \right)  &=&-i\left[ g_{k}\left( \eta \right)
a_{k}\left( \eta _{0}\right) -g_{k}^{\ast }\left( \eta \right)
a_{-k}^{\dagger }\left( \eta _{0}\right) \right]~~ ,
\end{eqnarray}
where
\begin{eqnarray}
f_{k}\left( \eta \right)  &=&\frac{1}{\sqrt{2k}}\left( u_{k}\left( \eta
\right) +v_{k}^{\ast }\left( \eta \right) \right)    \\
g_{k}\left( \eta \right)  &=&\sqrt{\frac{k}{2}}\left( u_{k}\left( \eta
\right) -v_{k}^{\ast }\left( \eta \right) \right)
\end{eqnarray}
and $f_{k}\left( \eta \right) $ is a solution of (\ref{deq1}). Following \cite{cite3} one can choose the vacuum to be
\begin{equation}
a_{k}\left( \eta _{0}\right) \left| 0,\eta _{0}\right\rangle =0~~.
\end{equation}
such that $v_{k}( \eta _{0}) =0$. It can also be shown that this vacuum is a zeroth-order adiabatic vacuum with a conjugate momenta given by
\begin{equation}
\pi _{k}\left( \eta _{0}\right) =ik\mu _{k}\left( \eta _{0}\right) ~~.
\end{equation}
If we now consider the standard treatment of fluctuation in inflation we can calculate the fluctuation spectrum as \cite{cite3}
\begin{equation}
{\cal P}_{\phi } =\left( \frac{H}{2\pi }\right) ^{2}\left[ 1+\left| \alpha _{k}\right|^{2}-\alpha _{k}e^{-2ik\eta _{0}}-\alpha _{k}^{\ast }e^{2ik\eta _{0}}\right] 
\frac{1}{1-\left| \alpha _{k}\right| ^{2}}~~,
\end{equation}
where $\alpha _{k}=\frac{i}{2k\eta _{0}+i}$. With $p=\Lambda $ where $\Lambda$ is the Plank scale considered here we get
\begin{equation}
\eta _{0}=-\frac{\Lambda }{Hk}~~.
\end{equation}
Assumption of $\frac{\Lambda }{H}\gg 1$ gives \cite{cite3}
\begin{equation}
\label{fluc}
{\cal P}_{\phi }=\left( \frac{H}{2\pi }\right) ^{2}\left[ 1-\frac{H}{\Lambda }\sin \left( \frac{2\Lambda }{H}\right) \right] ~~.
\end{equation}
It is interesting to note $\eta_0$ depends on $k$. Here the initial condition is not imposed in the infinite past but at the time when a particular mode is of the order of Plank scale. When $H$ is slowly varying, the spectrum is not scale invariant and there will be a modulation of ${\cal P}_{\phi}$ through the dependence of $H$ on the particular mode.
\par 
Now we can study the effect of MUP in the context of inflation. At first let us concentrate on the Hubble parameter. Using equation (\ref{g2}) we can modify the co-moving momentum $k$ to $k[1-\beta k+ 2\beta^2 k^2]$. Here we have made a change in notation for equation (\ref{g2}) and have used $\beta =l$. This modification is also studied in \cite{cite1,cite2,cite4} where the inflation scenario is investigated by the GUP as introduced in \cite{kem}. At the horizon crossing epoch we have \cite{lidsey,cite6}
\begin{equation}
\frac{dH}{dk}=-\frac{\epsilon_V H}{k}~~.
\end{equation}
With our MUP we can easily find
\begin{equation}
\label{eqn26}
H_M = \left[\frac{k}{\sqrt{1-\beta k +2\beta^2 k^2}}\right]^{-\epsilon_V} ~ e^{-\frac{\epsilon_V}{\sqrt{7}}~tan^{-1}\left(\frac{4\beta k-1}{\sqrt{7}}\right)}~~.
\end{equation}
The subscript $M$ is for MUP. Following equation (\ref{fluc}) we calculate the fluctuation spectrum as
\begin{equation}
{\cal P}_{\phi M}=\left( \frac{H_M}{2\pi }\right) ^{2}\left[ 1-\frac{H_M}{\Lambda }\sin \left( \frac{2\Lambda }{H_M}\right) \right] ~~.
\end{equation}
\begin{figure}[htb]
\begin{tabular}{c}
\includegraphics[width=7.5cm,height=4.5cm]{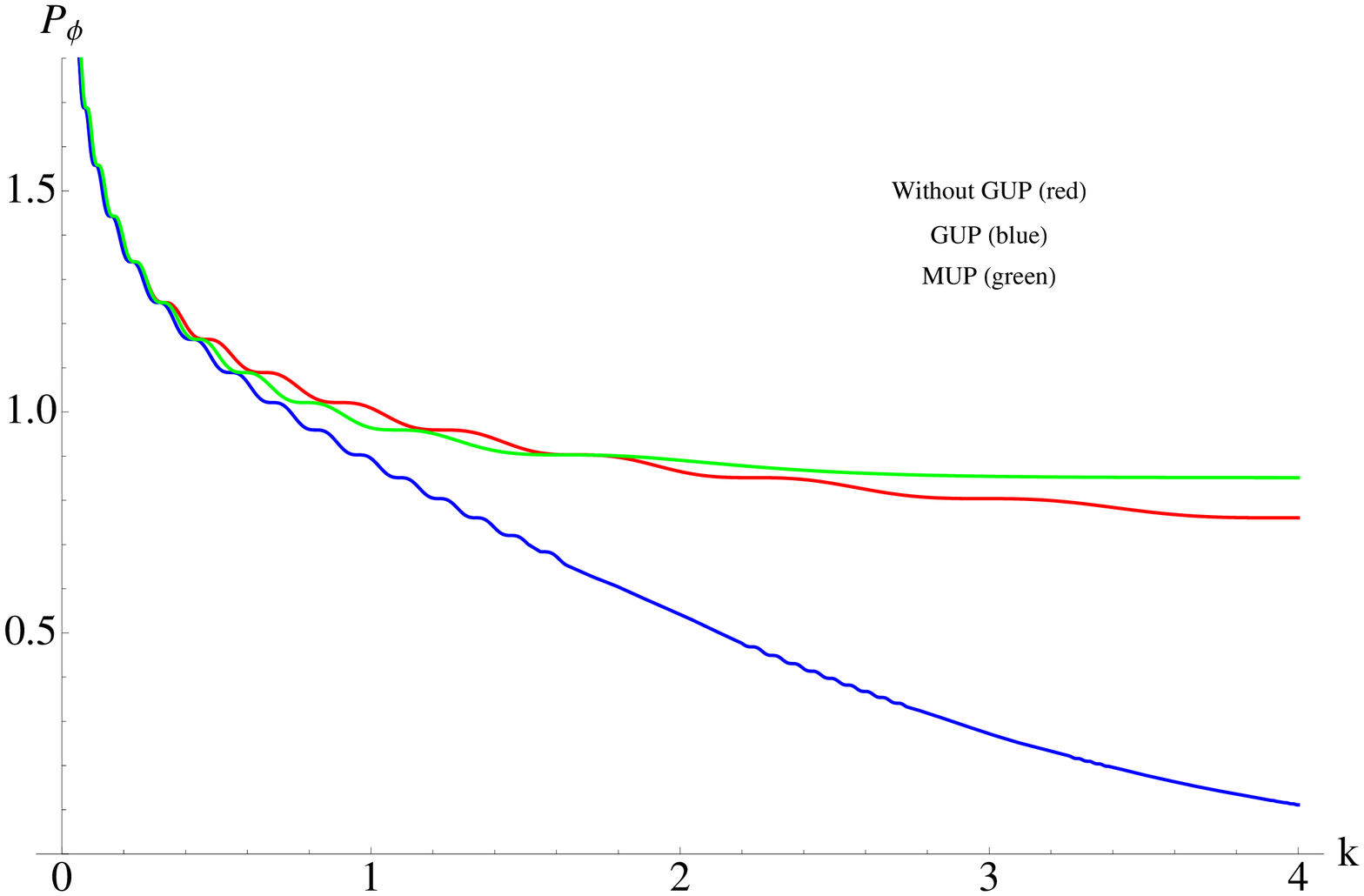} \hspace{.3cm}
\includegraphics[width=7.5cm,height=4.5cm]{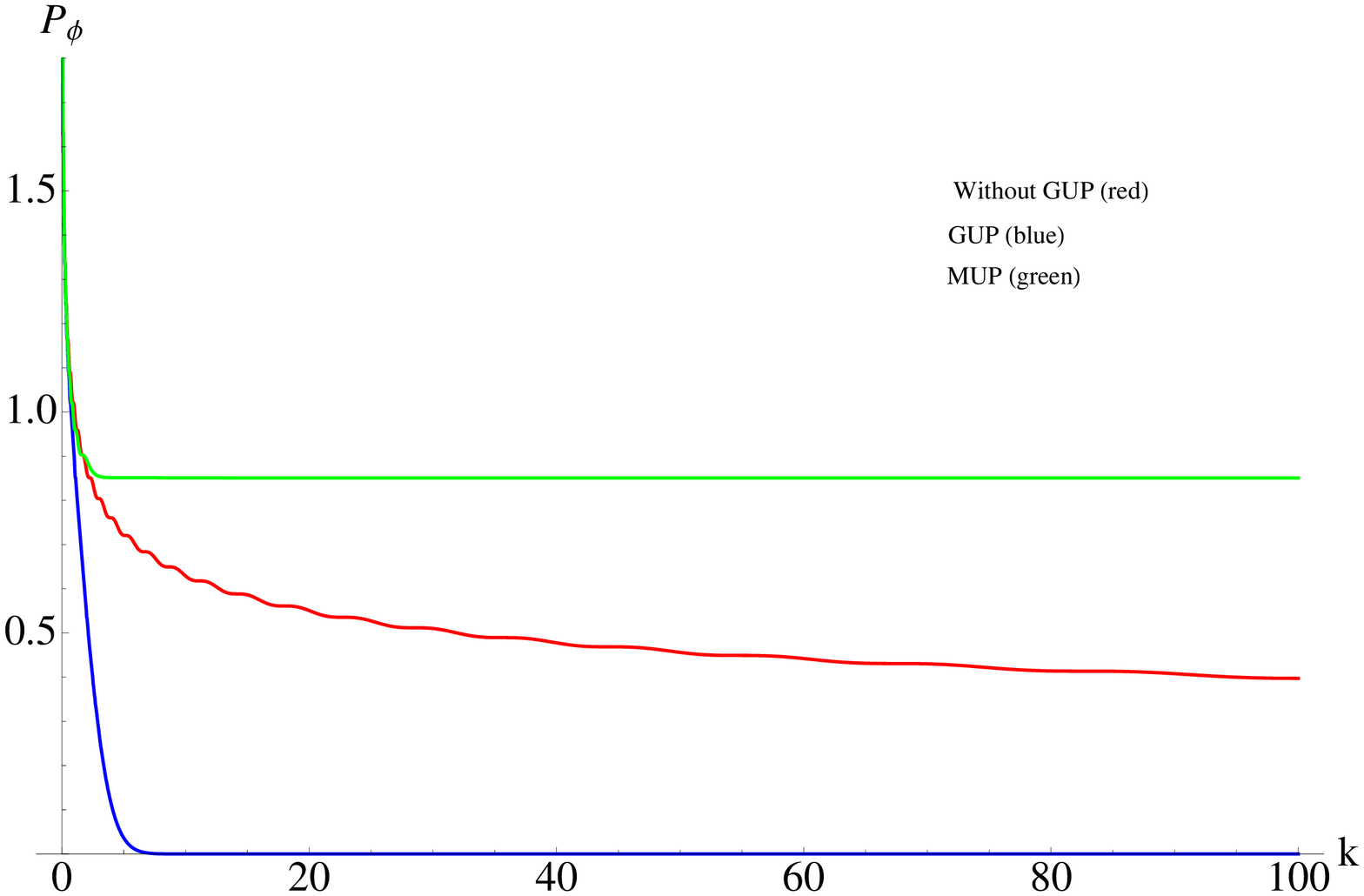}
\end{tabular}
\caption{\footnotesize  \textbf{(a)} Small $k$ dependence of the fluctuations with $\Lambda=100$, $\beta=.55$ and $\epsilon_V=.1$. \textbf{(b)} Large $k$ dependence of the fluctuations with $\Lambda=100$, $\beta=.55$ and $\epsilon_V=.1$. Red indicates the plot in Heisenberg setup, blue for the GUP \cite{cite4} and green for our MUP. A suitable scaling is used to enlarge the figure.}
\label{fig1ab}
\end{figure}
In Figure[\ref{fig1ab}] we have shown the dependence of $k$ on the fluctuation for fixed values of $\beta$ and $\epsilon_V$. [\ref{fig1ab}(a)] is for the small values of $k$ and [\ref{fig1ab}(b)] for the large values of $k$. We can see that there is an oscillatory behaviour in the $k$ dependence of the fluctuations. We can also see that the results found in \cite{cite4} with the GUP \cite{kem} shows great deviation from the Heisenberg setup and the fluctuations decay rapidly for large values of $k$. But in our case the deviation is much small and attains a constant value (independent of $k$) at large $k$ values. We expect that this behaviour may be observed in the CMB spectrum as a footprint of the strong effects of gravity in the early universe.
\par
\begin{figure}[htb]
\begin{tabular}{c}
\hspace{4cm} \includegraphics[width=8cm,height=5cm]{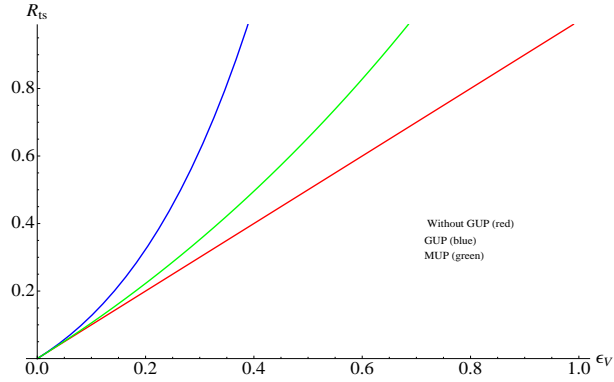}
\end{tabular}
\caption{\footnotesize  Plot of tensor-to-scalar ratio with the slow roll parameter $\epsilon_V $. Red indicates the plot in Heisenberg setup, blue for the GUP \cite{cite4} and green for our MUP. Here we have taken $\sqrt{16\pi} V=3 M_4$, $\beta=.55$ and $k=1$. A suitable scaling is used to enlarge the figure.}
\label{figratio}
\end{figure}

 
Here we have considered the field as a gravitational mode so ${\cal P}_{\phi}$ is actually the density fluctuation. For the scalar field a factor of $\left(\frac{H}{\dot \phi}\right)^2$ should be taken into account \cite{cite6}. So the tensor-to-scalar ratio ($R_{ts}$) is given by
\begin{align}
R_{ts} &= \left(\frac{\dot \phi_M}{H_M}\right)^2 \nonumber \\
& = \left(\frac{\sqrt{16\pi} V}{3 M_4}\right)^2 \left[\frac{\sqrt{\epsilon_V}}{\left(\frac{k}{\sqrt{1-\beta k + 2\beta^2 k^2}}\right)^{-2 \epsilon_V}}~~
e^{\frac{2 \epsilon_V}{\sqrt{7}}\tan ^{-1}\left(\frac{4\beta k -1}{\sqrt{7}}\right)}\right]^2 ~~,
\end{align}
where the subscript $M$ is for the MUP and $\dot \phi_M$ and $H_M$ respect the conditions $\frac{1}{2}\dot{\phi_M}^2 \ll V(\phi)$ and $3H_M\dot{\phi_M} \simeq -V'(\phi)$. Here $H_M$ is given by (\ref{eqn26}). In Figure[\ref{figratio}] we have shown the dependence of the tensor-to-scalar ratio on the inflation parameter $\epsilon_V$ for a particular value of $k$. We can see that the ratio increases if we take into account the quantum effects of gravity. The departures for GUP and MUP with respect to the standard setup is also seen in the plot.
\par
We now move on to study the effect of the MUP on the Randall-Sundrum type II inflationary scenario. Recently substantial amount of progress has been made in the field of brane-world cosmology. The Randall-Sundrum type II scenario \cite{cite7} has been studied much in detail where a single positive-tension brane is embedded in a five-dimensional space-time (termed as bulk) with a negative cosmological constant. This model has a non-trivial geometry in the bulk and it successfully gives four-dimensional general relativity in the low energy limit. For inflation one generally considers the dynamics of a scalar field on the five-dimensional bulk. The projection of the bulk inflaton on the brane behaves like an ordinary inflation in low energy regime. For a slowly varying potential of a scalar field there arises an effective non-zero cosmological constant on the brane and the brane may undergo inflation solely due to the dynamics in the bulk \cite{cite8,cite8a}.
\par
For the randall-Sundrum type II model the Friedmann equation can be written as \cite{cite9}
\begin{equation}
H^{2}=\frac{8\pi}{3M^{2}_{4}}~\rho\left[1+\frac{\rho}{2\lambda}\right] ~~,
\end{equation}
where $\lambda$ is the tension of the brane. While writing this equation we have neglected two terms in the right side. One with the cosmological constant and the other term often called as the dark radiation. The term comprising of the dark radiation originated as an integration constant from the projection of the bulk Weyl tensor transmitting bulk graviton influence onto the brane and it is assumed that it vanishes during inflation. A scalar field is the only matter here whose dynamics is considered, so the energy density $\rho$ is given by
\begin{equation}
 \rho=\frac{1}{2}\dot{\phi}^{2}+V(\phi)~~,
\end{equation}
where $V(\phi)$ is the inflation potential. In this model we can define two slow-roll parameters \cite{cite8}
\begin{equation}
\epsilon_V = \frac{M^{2}_{4}}{16\pi}\bigg(\frac{V'}{V} \bigg)^{2}\bigg[\frac{2\lambda(2\lambda+2V)}{(2\lambda+V)^{2}}\bigg]
\end{equation}
and
\begin{equation}
\eta_V = \frac{M^{2}_{4}}{8\pi}\bigg(\frac{V''}{V}\bigg)\bigg[\frac{2\lambda}{2\lambda+V}\bigg]~~.
\end{equation}
The scalar density perturbation in this model is given by \cite{cite6}
\begin{equation}
{\cal P}_{s(without MUP)} = \frac{9}{25} \frac{H^6}{V'^2} ~~.
\end{equation}
We now investigate the effect of MUP on the simplest chaotic inflation model driven by a scalar field with potential $V=\frac{1}{2}m^2 \phi ^2$. With this form of the potential the slow-roll parameter $\epsilon_V$ can be written as
\begin{equation}
\epsilon_V = \frac{2 M_4 \lambda}{\pi \phi^2} \left[\frac{2 \lambda + m^2 \phi^2}{(4 \lambda + m^2 \phi^2)^2}\right]~~.
\end{equation}
As a consequence of the MUP the scalar density perturbation ${\cal P}_{s(without MUP)}$ gets modified and it can now be written as
\begin{equation}
{\cal P}_s = \frac{9}{25 m^4 \phi^2} \left[
\left(\frac{k}{\sqrt{1-\beta k + 2\beta^2 k^2}}\right)^{-\frac{12 M_4 \lambda}{\pi \phi^2} \left[\frac{2 \lambda + m^2 \phi^2}{(4 \lambda + m^2 \phi^2)^2}\right]}~
e^{-\frac{12}{\sqrt{7}}  \frac{M_4 \lambda}{\pi \phi^2} \left[\frac{2 \lambda + m^2 \phi^2}{(4 \lambda + m^2 \phi^2)^2}\right] \tan^{-1} \left(\frac{4\beta k-1}{\sqrt{7}}\right)}\right]~~.
\end{equation}

\begin{figure}[htb]
\begin{tabular}{c}
\hspace{4cm} \includegraphics[width=8cm,height=5cm]{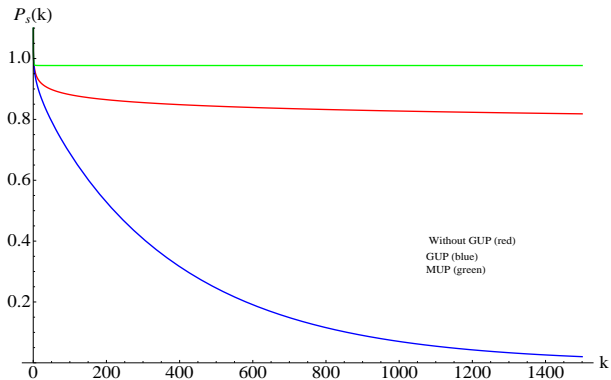}
\end{tabular}
\caption{\footnotesize  $k$ dependence of the scalar density fluctuation in the brane-world scenario. Red indicates the plot in Heisenberg setup, blue for the GUP \cite{cite4} and green for our MUP. Here we have taken $M_4=1$, $m=.1$, $\phi =4$, $\lambda =.1$ and $\beta=.55$. A suitable scaling is used to enlarge the figure.}
\label{figbrane}
\end{figure}
In Figure[\ref{figbrane}] we have shown the $k$ dependence of the scalar density perturbation. We can clearly see a distinct difference between the effect of the GUP and the MUP. In the MUP unlike the GUP we have a term in the commutation relation which is linear in Plank length (equation (\ref{g2})) as $ \beta= l=\frac{l_0 l_{pl}}{\hbar} $. Only due to this term we can see that ${\cal P}_s$ grabs a constant value (which is higher than the standard setup) and becomes independent of $k$ for large values of $k$. But as an effect of the GUP \cite{cite4} we have seen that ${\cal P}_s$ reduces and the value deviates much more for higher values of $k$. 
\par
So in this Letter we have studied the effects of the Modified Uncertainty Principle as proposed in \cite{b1} on the inflationary dynamics of the early universe in both standard and Randall-Sundrum type II scenarios. In the first place we have noticed that the Hubble parameter gets modified under the incorporation of the strong quantum effect of gravity. Later we studied the $k$ dependence of density fluctuation and found an oscillatory behaviour for small values of $k$. For large values of $k$ the density fluctuation approximately attains a constant value (nearly independent of $k$) which is slightly higher compared to the standard Heisenberg setup. But here we see a distinct difference if we compare our result with those found as an effect of the Generalized Uncertainty Principle as proposed in \cite{kem}. There we see the amplitude of density fluctuation is reduced by the incorporation of the GUP. But still we can argue that any deviation from the standard result contains a footprint of the quantum effect of gravity and a manifestation of these effects may be traced in the CMB spectrum. We also found that the tensor-to-scalar ratio increases by the incorporation of our Modified Uncertainty Principle. Later we studied the same for the Randall-Sundrum type II scenario and we found that the scalar density perturbation also gets modified. Here also we found that for large values of $k$ the scalar density fluctuation approximately attains a constant value (nearly independent of $k$) which is slightly higher compared to the standard Heisenberg setup. Finally we may conclude that the strong quantum effects of gravity (which may be incorporated through the Modified Uncertainty Principle) may exactly explain the observed spectrum of the Cosmic Microwave Background Radiation which in turn is an indirect test of the proposal itself.
 
\section*{Acknowledgements}
The author is very much thankful to Prof. Narayan Banerjee for helpful discussions and guidance. The author would also like to thank an anonymous referee for helpful comments and enlightening suggestions which immensely helped me improve the manuscript.

\end{document}